\documentclass[aps,prb,showpacs,twocolumn]{revtex4}
\usepackage{mathrsfs}
\usepackage{amssymb}
\usepackage{amsmath}
\usepackage{bm}
\usepackage{graphicx}
\usepackage{natbib}

\begin{document}
\title{Influence of an embedded quantum dot on the Josephson effect in the topological superconducting junction with Majorana doublets}
\author{Wei-Jiang Gong}
\email[Email address: ] {gwj@mail.neu.edu.cn}
\author{Zhen Gao}
\author{Wan-Fei Shan}
\author{Guang-Yu Yi}
\affiliation{
1. College of Sciences, Northeastern University, Shenyang 110819, China}
\date{\today}

\begin{abstract}
One Majorana doublet can be realized at each end of the time-reversal-invariant Majorana nanowires. We investigate the Josephson effect in the Majorana-doublet-presented junction modified by different inter-doublet coupling manners. It is found that when the Majorana doublets couple indirectly via a non-magnetic quantum dot, only the normal Josephson effects occur, and the fermion parity in the system just affects the current direction and amplitude. However, in the odd-parity case, applying finite magnetic field on the quantum dot can induce the appearance of the fractional Josephson effect. Next, when the direct and indirect couplings between the Majorana doublets coexist, no fractional Josephson effect takes place, regardless of finite magnetic field on the quantum dot. Instead, the $\pi$-period current has an opportunity to appear in some special cases. All the results are clarified by analyzing the influence of the fermion occupation in the quantum dot on the parity conservation in the whole system. We ascertain that this work will be helpful for describing the dot-assisted Josephson effect between the Majorana doublets.
\end{abstract}

 \maketitle

\bigskip

\section{Introduction}
Topological superconductor (TS) has received considerable experimental and theoretical attentions because Majorana zero-energy modes appear at the ends of the one-dimensional TS which can potentially be used for decoherence-free quantum computation.\cite{Majorana1,RMP1,Zhang1,Zhang2} In comparison with the conventional superconductor, the TS system shows new and interesting properties.\cite{Majorana2} For instance, in the proximity-coupled semiconductor-TS devices, the Majorana zero modes induce the zero-bias anomaly.\cite{Fuliang} A more compelling TS signature is the unusual Josephson current-phase relation. Namely, when the normal s-wave superconductor nano-wire is replaced by a TS wire with the Majorana zero modes, the current-phase relation will be modified to be $I_J \sim \sin{\phi\over2}$ and the period of the Josephson current vs $\phi$ will be $4\pi$ ($\phi$ is the superconducting phase difference). This is the so-called the fractional Josephson effect.\cite{Josephson1,Josephson2,Josephson3} Such a result can be understood in terms of fermion parity (FP). If the FP is preserved, there will be a protected crossing of the Majorana bound states at $\phi=\pi$ with perfect population inversion. As a result, the system cannot remain in the ground state as $\phi$ evolves from $0$ to $2\pi$ adiabatically.\cite{Aguado}
\par
Recently, a new class of topological superconductors with time-reversal symmetry, referred to as a DIII symmetry-class superconductor and classified by the $Z_2$ topological invariant,\cite{Ryu,Qi,Teo,Timm,Beenakker2} has attracted rapidly growing efforts.\cite{Deng,Nakosai,Wong,Zhang,Nagaosa2} Differently from chiral superconductors, in
DIII-class superconductors, the zero modes come in pairs due to Kramers's theorem. Up to now, many schemes have been proposed to realize $Z_2$ time-reversal-invariant Majorana nanowires using the proximity effects of $d$-wave, $p$-wave, $s\pm$-wave, or conventional $s$-wave superconductors.\cite{Stern,Flensberg,Haim,ZhangF2,Sau,Loss} It was shown that at each end of such a nanowire are localized two Majorana bound states that form a Kramers doublet and are protected by time-reversal symmetry. It is certain that the Majorana doublet can induce some new and interesting transport behaviors. Some groups have reported that in the Josephson junction formed by the Majorana doublet, the Josephson currents  show different periods in the cases of different FPs.\cite{Liuxj} However, for completely describing the transport properties of the Majorana doublet, any new proposals are desirable.
\par
In this work, we aim to investigate the influence of an embedded quantum dot (QD) on the current properties in the Josephson junction contributed by the Majorana doublets. Our motivation is based on the following two aspects. First, QD is able to accommodate an electron and the average electron occupation in one QD can be changed via shifting the QD level. Thus, when one QD is introduced in the TS junction, the FP can be re-regulated and the fractional Josephson current can be modified. Moreover, some special QD geometries can induce the typical quantum interference mechanisms, e.g., the Fano interference,\cite{Fanormp} which are certain to play an important role in adjusting the fractional Josephson effect. Second, one QD can mimic a quantum impurity in the practical system, which is able to provide some useful information for relevant experiments. Our calculations show that when the Majorana doublets couple indirectly via a non-magnetic QD, only the normal Josephson effects take place, irrelevant to the change of the FP. When a finite magnetic field is applied on the QD, the fractional Josephson effect comes into being in the odd-FP case. On the other hand, when the direct and indirect couplings between the Majorana doublets coexist, no fractional Josepshon effect occurs despite the presence of magnetic field on the QD, and the periods of the Josephson currents have opportunities to present the $\pi$-to-$2\pi$ transition when the QD level is shifted. The results in this work will be helpful for describing the QD-assisted Josephson effect between the Majorana doublets.

\begin{figure}
\centering\scalebox{0.45}{\includegraphics{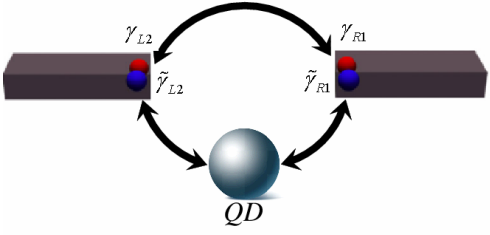}}
\caption{The Josephson junction formed by the direct coupling between the Majorana doublets and their
indirect coupling via a QD.} \label{Struct}
\end{figure}
\section{model\label{theory}}
The Josephson junction that we consider is formed by the direct couplings between the Majorana nanowires and their indirect couplings via a QD, as illustrated in Fig.1. The particle
tunneling process in this junction can be described by Hamiltonian $H_T$ with
\begin{equation}
H_T=\sum_{\alpha=L,R} H_{\alpha M}+H_{T0}+H_{TI}.
\end{equation}
$H_{\alpha M}$ denotes the particle motion in the two Majorana nanowires. With the proximity-induced $p$-wave and $s$-wave superconducting pairs, the effective tight-binding Hamiltonian in the $\alpha$-th nanowire can be written as\cite{Liuxj}
\begin{eqnarray}
&&H_{\alpha M}\notag\\
&&=\sum_{j\sigma}t_{\alpha j}c^\dag_{\alpha,j\sigma}c_{\alpha,j+1\sigma}+\sum_{j}(t_{\alpha,so}c^\dag_{\alpha,j\uparrow}c_{\alpha,j+1\downarrow}+H.c.)\notag\\
&&+\sum_{j}(\Delta_{\alpha p}c^\dag_{\alpha,j\uparrow}c^\dag_{\alpha,j+1\uparrow}+\Delta^*_{\alpha p}c^\dag_{\alpha,j\downarrow}c^\dag_{\alpha,j+1\downarrow}+H.c.)\notag\\
&&+\sum_{j}(\Delta_{\alpha s}c^\dag_{\alpha,j\uparrow}c^\dag_{\alpha,j\downarrow}+H.c.)-\mu_\alpha\sum_{j\sigma}n_{j\sigma}.
\end{eqnarray}
$c^\dag_{\alpha,j\sigma}$ and $c_{\alpha,j\sigma}$ ($\sigma=\uparrow,\downarrow$ or $\pm1$) are the electron creation and annihilation operators for the $j$-th site in the $\alpha$-th
nanowire. $t_{j\alpha}$ is the inter-site hopping energy and $t_{\alpha,so}$ represents the strength of spin-orbit coupling. $\Delta_{\alpha p}$ and $\Delta_{\alpha s}$ denote the energies of the $p$-wave and $s$-wave superconducting pairings, respectively. $\mu_\alpha$ is the chemical potential in the $\alpha$-th nanowire. Note that the hopping coefficients and the chemical potential are generically reonormalized by the proximity effect. The second term $H_{T0}$ denotes the direct coupling between the two Majorana nanowires, which can be expressed as
\begin{equation}
H_{T0}=\sum_\sigma\Upsilon c^\dag_{L,N\sigma}c_{R,1\sigma}+H.c.,   \label{Hamilt1}
\end{equation}
where $\Upsilon$ is the
direct coupling coefficient. Next, $H_{TI}$ is to express the indirect coupling between the two Majorana nanowires due to the presence of an embedded  QD (or a quantum impurity). Its expression can be given by
\begin{eqnarray}
H_{TI}&=&\sum_{\sigma}\varepsilon_0 d^\dag_\sigma d_\sigma+R(d^\dag_\uparrow d_\downarrow+d^\dag_\downarrow d_\uparrow)+Un_{d\uparrow}n_{d\downarrow}\notag\\
&&+\sum_\sigma V_Ld^\dag_\sigma c_{L,N\sigma}+\sum_\sigma V_Rd^\dag_\sigma c_{R,1\sigma}+H.c..   \label{Hamilt2}
\end{eqnarray}
Here $d^\dag_\sigma$ and $d_\sigma$ are the electron creation and annihilation operators in the QD, and $\varepsilon_0$ is the QD level. $R$ denotes the strength of an effective magnetic field applied on the QD, and $U$ denotes the intradot electron interaction with $n_{d\sigma}=d^\dag_\sigma d_\sigma$. In addition, $V_\alpha$ is coupling coefficient between the QD and the $\alpha$-th Majorana nanowire.
\par
In order to discuss the Josephson effect in this junction, we have to deduce an effective Hamiltonian that reflects the
direct and indirect couplings between the Majorana doublets. For this purpose, we define the Majorana operators
\begin{eqnarray}
&&\gamma_{\alpha1}=\sum_j[\mu^{(1)}_{\alpha j}c_{\alpha j}+\mu^{(1)*}_{\alpha j}c^\dag_{\alpha j}], \notag\\
&&\tilde{\gamma}_{\alpha1}=\sum_j[\tilde{\mu}^{(1)}_{\alpha j}\tilde{c}_{\alpha j}+\tilde{\mu}^{(1)*}_{\alpha j}\tilde{c}^\dag_{\alpha j}], \notag\\
&&\gamma_{\alpha2}=\sum_j[\mu^{(2)}_{\alpha j}c_{Lj}+\mu^{(2)*}_{\alpha j}c^\dag_{\alpha j}],\notag\\
&&\tilde{\gamma}_{\alpha2}=\sum_j[\tilde{\mu}^{(2)}_{\alpha j}\tilde{c}_{\alpha j}+\tilde{\mu}^{(2)*}_{\alpha j}\tilde{c}^\dag_{\alpha j}],
\end{eqnarray}
where $c_{\alpha j}=s_{1}c_{\alpha,j\uparrow}+s_{2}c_{\alpha,j\downarrow}$ is the renormalized electron operator at the $j$-th site in the $\alpha$-th site with $\tilde{c}_{\alpha j}={\cal T}c_{\alpha j}{\cal T}^{-1}$. Using the above formulas, we can solve the electron operators in terms of Majorana and nonzero-energy quasiparticle operators. Reexpressing the quasiparticles in terms of electron operators, we can interpret $c_{LN}$, $\tilde{c}_{LN}$, $c_{R1}$, and $\tilde{c}_{R1}$ by
\begin{eqnarray}
c_{LN}&=&\mu^{(2)*}_{LN}\gamma_{L2}-\sum_j a_{Lj}c_{Lj}-\sum_j b^*_{Lj}c^\dag_{Lj},\notag\\
\tilde{c}_{LN}&=&\tilde{\mu}^{(2)*}_{LN}\tilde{\gamma}_{L2}-\sum_j \tilde{a}_{Lj}\tilde{c}_{Lj}-\sum_j \tilde{b}^*_{Lj}\tilde{c}^\dag_{Lj},\notag\\
c_{R1}&=&\mu^{(1)*}_{R1}\gamma_{R1}-\sum_j a_{Rj}c_{Rj}-\sum_j b^*_{Rj}c^\dag_{Rj},\notag\\
\tilde{c}_{R1}&=&\tilde{\mu}^{(1)*}_{R1}\tilde{\gamma}_{R1}-\sum_j \tilde{a}_{Rj}\tilde{c}_{Rj}-\sum_j \tilde{b}^*_{Rj}\tilde{c}^\dag_{Rj},\label{Transform}
\end{eqnarray}
in which the normalization factor has been neglected. Besides, $a_{\alpha j}$, $\tilde{a}_{\alpha j}$ and $b_{\alpha j}$, $\tilde{b}_{\alpha j}$ are expansion coefficients, originated from the quasiparticle operators other than the corresponding Majorana mode.
Substituting Eq.(\ref{Transform}) into the expression of $H_T$, we can obtain the low-energy effective Hamiltonian of $H_T$ in the case of infinitely-long nanowires, which is divided into two parts. The first part is
\begin{eqnarray}
&&{\cal H}^{(0)}_T=i\Gamma_0\cos{\phi\over2}(\gamma_{L2}\gamma_{R1}-\tilde{\gamma}_{L2}\tilde{\gamma}_{R1})+\varepsilon_s d^\dag_s d_s\notag\\
&&+Un_{s}n_{\tilde{s}}-iW_Le^{-i\phi/2}d^\dag_s\gamma_{L2}+iW_Le^{-i\phi/2}d^\dag_{\tilde{s}}\tilde{\gamma}_{L2}\notag\\
&&+\varepsilon_{\tilde{s}}d^\dag_{\tilde{s}} d_{\tilde{s}}+W_Rd^\dag_s\gamma_{R1}+W_Rd^\dag_{\tilde{s}}\tilde{\gamma}_{R1}+H.c..
\end{eqnarray}
The relevant parameters here are defined as follows:
$\Gamma_0=2\Upsilon|\mu^{(2)}_{LN}\mu^{(1)}_{R1}|$, $W_L=V_L|\mu^{(2)}_{LN}|$, and $W_R=V_R|\mu^{(1)}_{R1}|$ in which $\mu^{(2)}_{LN}=i|\mu^{(2)}_{LN}|e^{i\phi_L/2}$, $\tilde{\mu}^{(2)}_{LN}=-i|\mu^{(2)}_{LN}|e^{i\phi_L/2}$, $\mu^{(1)}_{R1}=|\mu_{R1}^{(1)}|e^{i\phi_R/2}$, and $\tilde{\mu}^{(1)}_{R1}=|\mu^{(1)}_{R1}|e^{i\phi_R/2}$ (It is reasonable to suppose $|\mu^{(2)}_{LN}|=|\tilde{\mu}^{(2)}_{LN}|$ and $|\mu^{(1)}_{R1}|=|\tilde{\mu}^{(1)}_{R1}|$). Besides, in the above formula
$d^\dag_s=(s^*_{1}d^\dag_\uparrow+s^*_{2}d^\dag_\downarrow)$, $d^\dag_{\tilde{s}}=(-s^*_{2}d^\dag_\uparrow+s^*_{1}d^\dag_\downarrow)$, and $\varepsilon_{s/\tilde{s}}=\varepsilon_0\pm R$ with $n_s=d^\dag_s d_s$ and $n_{\tilde{s}}=d^\dag_{\tilde{s}} d_{\tilde{s}}$.
\par
For the second part, when the highest-order terms are neglected, it can be approximated as
\begin{eqnarray}
&&{\cal H}^{(1)}_T=-\Upsilon\mu^{(2)}_{LN}\gamma_{L2}(\sum_j a_{Rj}c_{Rj}+\sum_j b^*_{Rj}c^\dag_{Rj})\notag\\
&&-\Upsilon\mu^{(1)}_{R1}\gamma_{R1}(\sum_j a_{Lj}c_{Lj}+\sum_j b^*_{Lj}c^\dag_{Lj})\notag\\
&&-\Upsilon\tilde{\mu}^{(2)}_{LN}\tilde{\gamma}_{L2}(\sum_j \tilde{a}_{Rj}\tilde{c}_{Rj}+\sum_j \tilde{b}^*_{Rj}\tilde{c}^\dag_{Rj})\notag\\
&&-\Upsilon\tilde{\mu}^{(1)}_{R1}\tilde{\gamma}_{R1}(\sum_j \tilde{a}_{Lj}\tilde{c}_{Lj}+\sum_j \tilde{b}^*_{Lj}\tilde{c}^\dag_{Lj})\notag\\
&&-\sum_\alpha V_\alpha d^\dag_s(\sum_j a_{\alpha j}c_{\alpha j}+\sum_j b^*_{\alpha j}c^\dag_{\alpha j})\notag\\
&&-\sum_\alpha V_\alpha d^\dag_{\tilde{s}}(\sum_j \tilde{a}_{\alpha j}\tilde{c}_{\alpha j}+\sum_j \tilde{b}^*_{\alpha j}\tilde{c}^\dag_{\alpha j})+H.c..\label{HHt1}
\end{eqnarray}
It should be noted that since the $s$-wave pairing is present in the quantum wires, the electrons $c_{\alpha}$ and $\tilde{c}_\alpha$ will form a Cooper pair and condense. This process leads to an effective coupling between Majorana zero modes localized at the same end and the finite coupling between the Kramers doublet in the QD.
Therefore, up to the second-order perturbation in the tunneling process, we can express ${\cal H}^{(1)}_T$ as
\begin{eqnarray}
&&{\cal H}^{(1)}_T={1\over2}(\Upsilon^2 \mu^{(2)}_{LN}\tilde{\mu}^{(2)}_{LN}\gamma_{L2}\tilde{\gamma}_{L2}+V^2_Rd^\dag_sd^\dag_{\tilde{s}}){\cal G}_R\notag\\
&&+{1\over2}(\Upsilon^2 \mu^{(1)}_{R1}\tilde{\mu}^{(1)}_{R1}\gamma_{R1}\tilde{\gamma}_{R1}+V^2_Ld^\dag_sd^\dag_{\tilde{s}}){\cal G}_L+H.c.,
\end{eqnarray}
where ${\cal G}_\alpha=\sum_ja_{\alpha j}\tilde{a}_{\alpha j}\int d\tau\langle T_\tau c_{\alpha j}(\tau)\tilde{c}_{\alpha j}(0)\rangle+\sum_jb^*_{\alpha j}\tilde{b}^*_{\alpha j}\int d\tau\langle T_\tau c^\dag_{\alpha j}(\tau)\tilde{c}^\dag_{\alpha j}(0)\rangle$ with $\int d\tau\langle\cdots\rangle$ being a time-ordered integral. In the case of uniform superconducting pairings in the
Majorana nanowires, ${\cal G}_\alpha$ can be further deduced as ${\cal G}_\alpha=\sum_{j}a_{\alpha j}\tilde{a}_{\alpha j}\sum_k{\Delta^*_{\alpha s}\over\xi^2_{\alpha k}(\Delta_{\alpha s},\Delta_{\alpha p})}
-\sum_jb^*_{\alpha j}\tilde{b}^*_{\alpha j}\sum_k{\Delta_{\alpha s}\over\xi^2_{\alpha k}(\Delta_{\alpha s},\Delta_{\alpha p})}$ in which $\xi_{\alpha k}$ is the eigen-energy of the isolated superconductor.\cite{Liuxj} With the relations in Eq.(\ref{Transform}), we can get the relationship that $a_{\alpha j}\tilde{a}_{\alpha j}=|a_{\alpha j}\tilde{a}_{\alpha j}|$ and $b_{\alpha j}\tilde{b}_{\alpha j}=|b_{\alpha j}\tilde{b}_{\alpha j}|e^{2i\phi_\alpha}$. Accordingly, ${\cal H}^{(1)}_T$ can be written as
\begin{eqnarray}
{\cal H}^{(1)}_T&=&i\Gamma_{1L}\sin\phi\gamma_{L2}\tilde{\gamma}_{L2}-i\Gamma_{1R}\sin\phi\gamma_{R1}\tilde{\gamma}_{R1}\notag\\
&&+(\Gamma_{2L}e^{-i\phi}+\Gamma_{2R})d^\dag_sd^\dag_{\tilde{s}}+H.c.,
\end{eqnarray}
in which $\Gamma_{1L}=\Upsilon^2|\mu^{(2)}_{LN}|^2|{\cal G}_R|$, $\Gamma_{1R}=\Upsilon^2|\mu^{(1)}_{R1}|^2|{\cal G}_L|$, and $\Gamma_{2\alpha}={1\over2}V^2_\alpha|{\cal G}_\alpha|$. Up to now, we have obtained the low-energy effective Hamiltonian of such a structure.
\par
The phase difference between the two Majorana wires will drive finite Josephson current through them, which can be directly evaluated by the following formula
\begin{equation}
I_J={2e\over \hbar}{\langle{\partial {\cal H}_{T}}\rangle\over\partial \phi}
\end{equation}
with $\langle\cdots\rangle$ being the thermal average. It is certain that solving the Josephson current is dependent on the diagonalization of ${\cal H}_T$.
\par
In the following, we try to diagonalize the Hamiltonian. First, by defining $\gamma_{1}={1\over\sqrt{2}}(\gamma_{L2}+\tilde{\gamma}_{R1})$ and $\gamma_{2}={1\over\sqrt{2}}(\gamma_{R1}+\tilde{\gamma}_{L2})$ with $\tilde{\gamma}_{j}={\cal T}\gamma_j{\cal T}^{-1}$,
we reexpress $H_T$ as
\begin{eqnarray}
&&{\cal H}_T=i(\Gamma_0\cos{\phi\over2}+\Gamma_1\sin\phi)\gamma_{1}\gamma_{2}+\varepsilon_sd^\dag_s d_s\notag\\
&&-i(\Gamma_0\cos{\phi\over2}-\Gamma_1\sin\phi)\tilde{\gamma}_{1}\tilde{\gamma}_{2}+\varepsilon_{\tilde{s}}d^\dag_{\tilde{s}} d_{\tilde{s}}\notag\\
&&+Un_{s}n_{\tilde{s}}+(\Gamma_{2L}e^{-i\phi}+\Gamma_{2R})d^\dag_sd^\dag_{\tilde{s}}\notag\\
&&+{1\over\sqrt{2}}(-iW_Le^{-i\phi/2}d^\dag_s+W_Rd^\dag_{\tilde{s}})\gamma_1\notag\\
&&+{1\over\sqrt{2}}(iW_Le^{-i\phi/2}d^\dag_{\tilde{s}}+W_Rd^\dag_{s})\gamma_2\notag\\
&&+{1\over\sqrt{2}}(iW_Le^{-i\phi/2}d^\dag_{\tilde{s}}-W_Rd^\dag_{s})\tilde{\gamma}_1\notag\\
&&+{1\over\sqrt{2}}(iW_Le^{-i\phi/2}d^\dag_s+W_Rd^\dag_{\tilde{s}})\tilde{\gamma}_2+H.c.,
\end{eqnarray}
where $\Gamma_{1\alpha}$ is supposed to be $\Gamma_1$. Next, ${\cal H}_T$ can be expressed in the normal fermion representation by supposing $\gamma_1=(f+f^\dag)$, $\gamma_2=i(f^\dag-f)$ and $\tilde{\gamma}_1=i(\tilde{f}^\dag-\tilde{f})$, $\tilde{\gamma}_2=(\tilde{f}+\tilde{f}^\dag)$ where $f^\dag$, $\tilde{f}^\dag$ and $f$, $\tilde{f}$ are the fermionic creation and annihilation
operators. Accordingly, the matrix form of ${\cal H}_T$ can be deduced on the basis of $|n_{s}n_{\tilde{s}}n_fn_{\tilde{f}}\rangle$ where $n_f=f^\dag f$ and $n_{\tilde{f}}=\tilde{f}^\dag \tilde{f}$. Note that in the system with Majorana bound states, only the FP is the good quantum number. Thus, we should build the Fock state according to the FP. First, in the case of the even FP, the Fock state can be written as  $|\Psi_e\rangle=a_1|0000\rangle+a_2|0011\rangle+a_3|0101\rangle+a_4|1001\rangle+a_5|0110\rangle+a_6|1010\rangle+a_7|1100\rangle+a_8|1111\rangle$. As a result, the matrix form of ${\cal H}^{(e)}_{T}$ can be written as

\begin{widetext}
\begin{eqnarray}
{\cal H}^{(e)}_{T}=\left[
\begin{array}{cccccccc}
-\Lambda &0&-{\cal A}&-i{\cal A}&-{\cal A}&i{\cal A}&{\cal D}&{\cal D}\\
 0&\Lambda &{\cal B}&i{\cal B}&-{\cal B}&i{\cal B}&{\cal D}&{\cal D}\\
 -{\cal A}^*&{\cal B}^*&\varepsilon_{\tilde{s}}-\Omega&0&0&0&i{\cal B}&-i{\cal A}\\
 i{\cal A}^*&-i{\cal B}^* &0&\varepsilon_s-\Omega&0&0&{\cal B}&-{\cal A}\\
 -{\cal A}^*&-{\cal B}^*&0&0&\varepsilon_{\tilde{s}}+\Omega&0&-i{\cal B}&-i{\cal A}\\
 -i{\cal A}^*&-i{\cal B}^*&0&0&0&\varepsilon_s+\Omega&{\cal B}&{\cal A}\\
 {\cal D}^*&{\cal D}^*&-i{\cal B}^*&{\cal B}^*&i{\cal B}^*&{\cal B}^*&\varepsilon_s+\varepsilon_{\tilde{s}}+U-\Lambda&0\\
 {\cal D}^*&{\cal D}^*&i{\cal A}^*&-{\cal A}^*&i{\cal A}^*&{\cal A}^*&0&\varepsilon_s+\varepsilon_{\tilde{s}}+U+\Lambda
\end{array}
\right],\label{matrix1}
\end{eqnarray}
\end{widetext}
where ${\cal A}=(W_Le^{i\phi/2}-W_R)/\sqrt{2}$, ${\cal B}=(W_Le^{i\phi/2}+W_R)/\sqrt{2}$, ${\cal D}=\Gamma_{2L}e^{i\phi}+\Gamma_{2R}$, $\Lambda=2\Gamma_0\cos{\phi\over2}$, and $\Omega=2\Gamma_1\sin\phi$. Next, for the case of the even FP, the Fock state can be written as  $|\Psi_o\rangle=b_1|0001\rangle+b_2|0010\rangle+b_3|0100\rangle+b_4|1000\rangle+b_5|0111\rangle+b_6|1011\rangle+b_7|1101\rangle+b_8|1110\rangle$ and the matrix of ${\cal H}^{(o)}_{T}$ takes the form as
 \begin{widetext}
\begin{eqnarray}
{\cal H}^{(o)}_{T}=\left[
\begin{array}{cccccccc}
-\Omega &0&{\cal B}&-i{\cal B}&-{\cal A}&i{\cal A}&{\cal D}&{\cal D}\\
 0&\Omega &{\cal B}&i{\cal B}&{\cal A}&i{\cal A}&{\cal D}&{\cal D}\\
 {\cal B}^*&{\cal B}^*&\varepsilon_{\tilde{s}}-\Lambda&0&0&0&i{\cal A}&-i{\cal A}\\
 i{\cal B}^*&-i{\cal B}^* &0&\varepsilon_s-\Lambda&0&0&-{\cal A}&-{\cal A}\\
 -{\cal A}^*&{\cal A}^*&0&0&\varepsilon_{\tilde{s}}+\Lambda&0&-i{\cal B}&-i{\cal B}\\
 -i{\cal A}^*&-i{\cal A}^*&0&0&0&\varepsilon_s+\Lambda&{\cal B}&-{\cal B}\\
{\cal D}^*&{\cal D}^*&-i{\cal A}^*&-{\cal A}^*&i{\cal B}^*&{\cal B}^*&\varepsilon_s+\varepsilon_{\tilde{s}}+U-\Omega&0\\
{\cal D}^*&{\cal D}^*&i{\cal A}^*&-{\cal A}^*&i{\cal B}^*&-{\cal B}^*&0&\varepsilon_s+\varepsilon_{\tilde{s}}+U+\Omega
\end{array}
\right].\label{matrix2}
\end{eqnarray}
\end{widetext}
\par
\par
For the extreme case of strong magnetic field limit, if $E_{\tilde{s}}$ is in the finite-energy region, $E_s$ will be empty, and then only one level contributes to the Josephson effects, respectively. Accordingly, in such a case, the matrixes of ${\cal H}^{(e)}_T$ and ${\cal H}^{(o)}_T$ will be halved, i.e.,
\begin{eqnarray}
{\cal H}^{(e)}_{T}=\left[
\begin{array}{cccc}
-\Lambda &0&-{\cal A}&-{\cal A}\\
 0&\Lambda &{\cal B}&-{\cal B}\\
 -{\cal A}^*&{\cal B}^* &\varepsilon_{\tilde{s}}-\Omega&0\\
 -{\cal A}^*&-{\cal B}^*&0&\varepsilon_{\tilde{s}}+\Omega
\end{array}
\right],\label{matrix5}
\end{eqnarray}
and
\begin{eqnarray}
{\cal H}^{(o)}_{T}=\left[
\begin{array}{cccc}
-\Omega &0&{\cal B}&-{\cal A}\\
 0&\Omega &{\cal B}&{\cal A}\\
 {\cal B}^*&{\cal B}^* &\varepsilon_{\tilde{s}}-\Lambda&0\\
 -{\cal A}^*&{\cal A}^*&0&\varepsilon_{\tilde{s}}+\Lambda
\end{array}
\right].\label{matrix6}
\end{eqnarray}
\par
Based on the above analysis, we can understand that at the zero-temperature limit, the Josephson current in this structure is dependent on the FP, i.e.,
\begin{equation}
I^{(e/o)}_J={2e\over \hbar}{\langle{\partial {\cal H}^{(e/o)}_{T}}\rangle\over\partial \phi}={2e\over \hbar}{{\partial E^{(e/o)}_{GS}}\over\partial \phi}.
\end{equation}
$E^{(e/o)}_{GS}$ are the ground-state (GS) energies in the even and odd FP cases, respectively.

\begin{figure}
\centering\scalebox{0.45}{\includegraphics{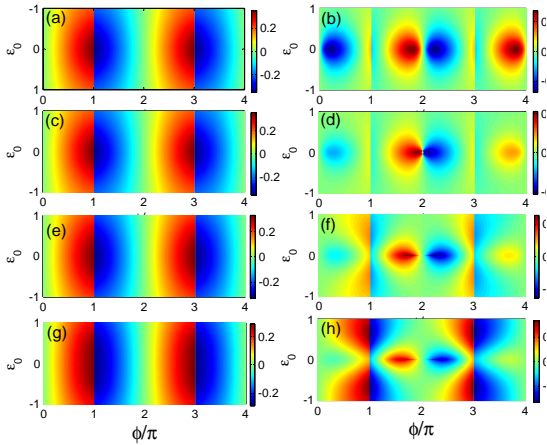}}
\caption{Josephson current spectra in the case where the Majorana doublets couple indirectly via a QD. The structural parameters are taken to be $W_\alpha=0.25$ and $\Gamma_{2\alpha}=0.05$. The left and right columns correspond to the even-FP and odd-FP results, respectively. (a)-(b) The case of the non-magnetic QD. (c)-(d) The case of finite magnetic field on the QD with $R=0.1$. (e)-(f) $R=0.3$. (g)-(h) $R=0.5$. } \label{Indirect1}
\end{figure}

\section{Numerical results and discussions \label{result2}}
Following the derivation in the above section, we next perform the numerical calculation to discuss the detailed properties of the Josephson current through such a system. As a typical case, only the zero-temperature limit is considered. Before calculation, we would like to review the Josephson effect in the case of $V_\alpha=0$.
In such a case, ${\cal H}_T=(\Gamma_0\cos{\phi\over2}+\Gamma_1\sin\phi)(2n_f-1)+(\Gamma_0\cos{\phi\over2}-\Gamma_1\sin\phi)(2n_{\tilde{f}}-1)$, and $|n_fn_{\tilde{f}}\rangle_f$ are the eigenstates of ${\cal H}_T$. The two even-FP eigenstates are $|00\rangle_f$ and $|11\rangle_f$, and their corresponding GS energies are $E^{(e)}_{GS}=\mp2\Gamma_0\cos{\phi\over2}$. Contrarily, the odd-FP eigenstates are $|10\rangle_f$ and $|01\rangle_f$ with the GS energies $E^{(o)}_{GS}=\pm2\Gamma_1{\sin\phi}$. Just as concluded in the previous works,\cite{Liuxj} the fractional Josephson effect occurs in the situation of even FP, otherwise only the normal Josephon effect can be observed.

\par
In Fig.2 we suppose $\Gamma_0=\Gamma_1=0$ and choose $W_\alpha=0.25$ and $\Gamma_{2\alpha}=0.05$ to investigate the Josephson effect in the case where the Majorana doublets couple indirectly to each other via a QD. The results are shown in Figs.2-3: Fig.2 corresponds to the noninteracting results, and Fig.3 describes the influences of the intradot Coulomb interaction on the Josephson effects in different FPs in the case of $U=2.0$. First, in Fig.2(a)-(b) we find that when a non-magnetic QD is presented, it induces the occurrence of the normal Josephson effects and the departure of $\varepsilon_0$ from zero weakens the current amplitudes, irrelevant to the FP difference. Also, the FP plays an important role in affecting the Josephson effects. Concretely, the Josephson currents in different FPs flow in the opposite directions for the same $\phi$, and the amplitude of $I^{(o)}_J$ is about one half of that of $I^{(e)}_J$ and when $|\varepsilon_0|>0.5$ $I^{(o)}_J$ gets close to zero. Besides, at the points of $\phi=(2n-1)\pi$, in the even-FP case the discontinuous change of the Josephson current is more well-defined compared with the odd-FP case. Next, when finite magnetic field is applied on the QD, the even-FP Josephson current shows little change except that its amplitude becomes less dependent on the QD-level shift. However, in the odd-FP case, the Josephson current changes completely. It can be clearly found that with the strengthening of the magnetic field, the original current oscillation is suppressed. Especially in the vicinity of $\phi=4n\pi$, the current amplitude tends to disappear at the case of $R=0.5$. Thus, it is certain that in the case of odd FP, a nonzero field on the QD can induce the occurrence of the fractional Josephson current. In addition, the increase of $R$ enhances the current oscillation around the points of $\phi=(2n-1)\pi$ when $\varepsilon_0$ departs from $\neq0$. Up to now, we can conclude that when the Majorana doublets are coupled by a magnetic QD, the fractional Josephson effect has an opportunity to take place, but it is different from the case where the Majorana doublets couple directly.\cite{Liuxj}
\begin{figure}
\centering\scalebox{0.46}{\includegraphics{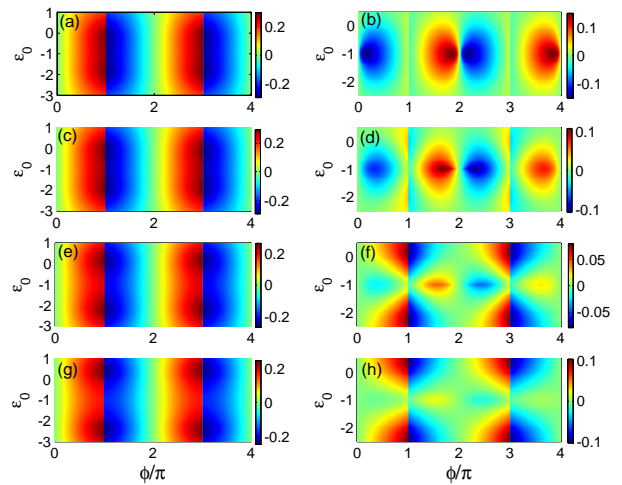}}
\caption{Josephson current in the case where the Majorana doublets couple indirectly via a finite-Coulomb QD. The Coulomb strength is $U=2.0$ and the others are the same as those in Fig.2. The left and right columns correspond to the even-FP and odd-FP results, respectively. (a)-(b) The case of the non-magnetic QD. (c)-(d) The case of finite magnetic field on the QD with $R=0.1$. (e)-(f) $R=0.3$. (g)-(h) $R=0.5$.} \label{Indirect2}
\end{figure}
\par
Coulomb interaction is a key factor to influence the characteristics of QD. In Fig.3 we consider the case of finite intradot Coulomb interaction and investigate the effect of the magnetic QD on the Josephson currents in the finite-Coulomb case. In Fig.3(a)-(b) we first find that in the even-FP case with a non-magnetic QD, the Coulomb interaction benefits the Josephson effect, since in the region of $-2.5<\varepsilon_0<0.5$ the current amplitude is relatively robust and weakly dependent on the shift of the QD level. In contrast, for the odd-FP case, the intradot Coulomb interaction only moves the current maximum to the point of $\varepsilon_0=-1.0$, but it does not vary the current oscillation manner compared with the noninteracting case. Hence, the Coulomb interaction only adjusts the effect of the QD level on the Josephson effects but does not modify the current oscillation manner with the change of $\phi$. Next, Fig.3(c)-(h) show that regardless of the FP difference, the current amplitudes are suppressed by the application of finite magnetic field on the QD. In the even-FP case, the current amplitude around the point of $\varepsilon_0=-1.0$ undergoes a relatively-apparent suppression. For the odd-FP case, except the suppression of the current amplitude, the fractional Josephson effect becomes weak but can still be observed.
\par
According to the results in the above paragraphes, when the Majorana doublets couple indirectly via a magnetic QD, the fractional Josephson effect has an opportunity to come into being in the odd-FP case. Otherwise, only the normal Josephson effect can be observed. In order to explain the results in Figs.2-3, we would like to compare the case of $\Upsilon\neq0$ and $V_\alpha=0$ with the case of $\Upsilon=0$ and $V_\alpha\neq0$. In the former case, the Josephson effects are only determined by the FP of state $|n_fn_{\tilde{f}}\rangle_f$. Namely, when the system is located at the states $|00\rangle_f$ or $|11\rangle_f$, the fractional Josephson effect occurs. However, when the Majorana doublets couple indirectly via a QD, the Fock space defined by $|n_fn_{\tilde{f}}\rangle_f$ just becomes a subspace of the Fock space formed by $|n_{s}n_{\tilde{s}}n_fn_{\tilde{f}}\rangle$. Since both $\Gamma_0$ and $\Gamma_1$ are equal to zero in such a case, the nonzero contributions of $|00\rangle_f$ and $|11\rangle_f$ to the Josephson effect are dependent on their finite indirect couplings, i.e., their simultaneous couplings to the other states. Thus, the manners of the indirect couplings inevitably regulate the properties of the Josephson effects. For instance, in Eq.(\ref{matrix1}) we find that in the even-FP case, the states that include $|00\rangle_f$ and $|11\rangle_f$ couple indirectly in the way of left-right symmetry. This leads to the equal contributions of $|00\rangle_f$ and $|11\rangle_f$ to the Josepshon effect, and then the normal Josephon effect takes place, independent of the presence of magnetic field. On the other hand, in the odd-FP case, Eq.(\ref{matrix1}) shows that the coupling between the states that include $|00\rangle_f$ and $|11\rangle_f$ is left-right asymmetric unless $\varepsilon_s\neq\varepsilon_{\tilde{s}}$. Consequently, if finite $R$ is considered, $|00\rangle_f$ and $|11\rangle_f$ make different contributions to the Josephson effect, which gives rise to the fractional Josephson effect. Also, note that in odd-FP case, the left-right asymmetric coupling manner weakens the quantum coherence and suppresses the current amplitude to some degree. When the Coulomb interaction is taken into account, the QD level splits. In the even-FP case, the picture in the noninteracting case can be doubled since the left-right symmetric coupling manner. Nevertheless, in the odd-FP case, only in the case of electron-hole symmetry, the left-right symmetric coupling manner can be satisfied. Therefore, the maximal Josephson current occurs at the point of $\varepsilon_0=-1.0$ in the case of $U=2.0$.
\begin{figure}
\centering\scalebox{0.43}{\includegraphics{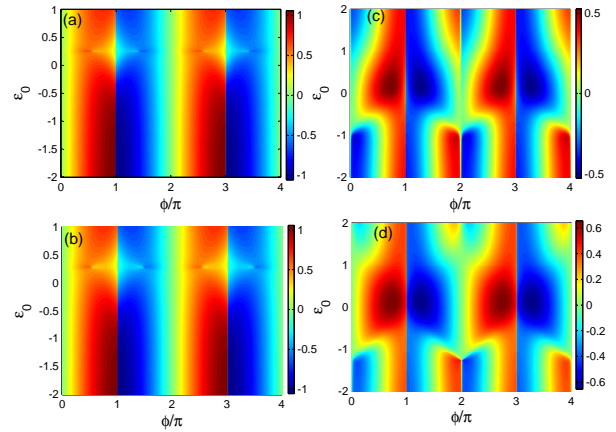}}
\caption{Josephson current spectra in the case where the direct and indirect couplings between the Majorana doublets co-exist. The relevant parameters are $\Gamma_0=0.5$, $\Gamma_1=0.1$, $W_\alpha=0.25$, and $\Gamma_{2\alpha}=0.05$. (a)-(b) The even-FP case with non-magnetic and magnetic QDs with $R=0.3$. (c)-(d) The odd-FP results with $R=0$ and $R=0.3$.} \label{Direct1}
\end{figure}

\par
We next proceed to pay attention to the Josephson effect in the case where the direct and indirect couplings between the Majorana doublets coexist. The results are shown in Figs.4-5 where $\Gamma_0$ is taken to be $0.5$ and $\Gamma_1=0.1$. The noninteracting results are presented in Fig.4 and Fig.5 describes the case of $U=2.0$. In Fig.4, we find that for any $\varepsilon_0$, the two kinds of Josephson currents show dissimilar oscillations with the adjustment of superconducting phase difference. In the even-FP case with $R=0$, when $\varepsilon_0$ gets approximately close to $0.25$, the amplitude of the Josephson current is decreased, otherwise, the Josephson effect will be enhanced and then holds. However, the current period always keeps to be $2\pi$, as shown in Fig.4(a). Next, in Fig.4(b) where $R=0.3$, we can see that the only effect of the magnetic field in the QD is to further suppress the minimum of the Josephson current. Such a result is exactly similar to the result of $\Gamma_0=0$. On the other hand, for the odd-FP case, in Fig.4(c), it shows that in the region of $\varepsilon_1<-1.0$, the Josephson current seems to oscillate in $\pi$ period. When the QD level increases to $\varepsilon_0=-1.0$, the current period experiences the $\pi$-to-$2\pi$ transition with the disappearance of the current around the points of $\phi=2n\pi$. Next, in the region of $-1.0<\varepsilon_0<1.0$, the Josephson current varies in $2\pi$ period with its maximum in the vicinity of $\varepsilon_0=0$. When $\varepsilon_0$ further increases from $1.0$, the Josephson current recovers the $\pi$-period oscillation gradually. Fig.4(d) presents the effect of the magnetic field on the QD in odd-FP case. It seems that in such a case, the magnetic field can not induce the fractional Josephson effect, but it tends to enhance the current amplitude in the region of $-1.0<\varepsilon_0<1.0$, which is exactly opposite to the case of $\Gamma_0=0$.

\begin{figure}
\centering\scalebox{0.43}{\includegraphics{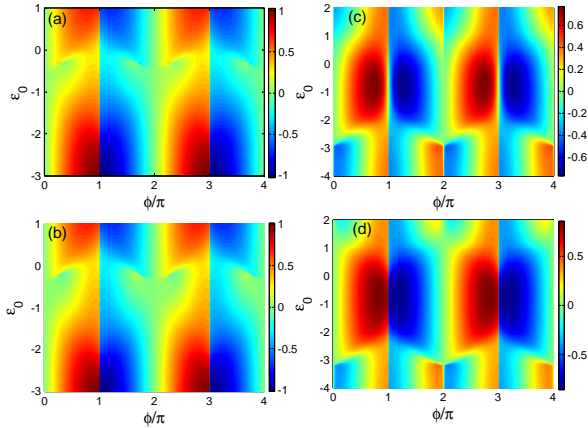}}
\caption{Josephson current spectra in the case of simultaneous direct and indirect couplings between the Majorana doublets. The Coulomb strength is fixed at $U=2.0$, and the other parameters are identical with those in Fig.4. (a)-(b) The even-FP case with non-magnetic and magnetic QDs with $R=0.3$. (c)-(d) The odd-FP results with $R=0$ and $R=0.3$.} \label{Direct2}
\end{figure}
\par
Following the above result, we present the influence of the magnetic field on the finite-Coulomb results, as displayed in Fig.5. Here, the Coulomb strength is also taken to be $U=2.0$. First, Fig.5(a) shows the even-FP result with the non-magnetic QD. We can find that in such a case, the current minimum is shifted to the position of $\varepsilon_0\approx-0.25$. Besides, the Coulomb interaction efficiently weakens the Josephson effect, since increasing $\varepsilon_0$ from $-2.0$ begins to eliminate the current amplitude gradually. Next when finite magnetic field is applied on the QD with $R=0.3$, it further suppresses the minimum of the Josephson current, similar to the noninteracting case [See Fig.5(b)]. The odd-FP results are shown in Fig.5(c)-(d), where the magnetic field strength is taken to be zero and $0.3$, respectively. In Fig.5(c), we see that different from the noninteracting result, the $2\pi$-period oscillation of the current occurs from $\varepsilon_0=-3.0$. In the region of $-3.0<\varepsilon_0<1.0$, the Josephson current varies in $2\pi$ period with its maximum in the vicinity of $\varepsilon_0=-1.0$. Besides, it can be noted that the Coulomb interaction enhance the amplitude of the Josephson current, in comparison with the noninteracting current results. For the effect of the magnetic field in the odd-FP case, as shown in Fig.5(d), it is analogous to that in the noninteracting case. Namely, it tends to enhance the amplitude of the $2\pi$-period current, despite in the region of $-3.0<\varepsilon_0<1.0$. Also, it does not induce the appearance of the fractional Josephson effect.
\par
The results in Figs.4-5 can be explained similar to the discussion about Figs.2-3. In the case of finite $\Upsilon$, the underlying physics that governs the Josephson effects becomes complicated. The reason arises from two aspects. First, the fermion number of the QD re-regulates the FP of $|n_fn_{\tilde{f}}\rangle_f$ for conserving the FP in the whole system. Second, the Fano interference can be induced by the direct and indirect couplings between the Majorana doublets. Based on this idea, we see that in the even-FP case with zero magnetic field, when the QD level is far away from the energy zero point, both $n_s$ and $n_{\tilde{s}}$ are equal to $1$ or $0$ simultaneously. Consequently, the states $|00\rangle_f$ and $|11\rangle_f$ contribute equally to the Josephson effect, so the normal Josephson effect occurs with the current amplitude proportional to $\Gamma_0$. Alternatively, in the odd-FP case with $|\varepsilon_{s/\tilde{s}}|\gg0$, the $\pi$-period current occurs and its amplitude is related to $\Gamma_1$, due to the co-contribution of states $|10\rangle_f$ and $|01\rangle_f$. When the QD level gets close to the energy zero point, it will be partly-occupied. In such a situation, $|10\rangle_f$ and $|01\rangle_f$ contribute to the even-FP Josephson current, whereas $|00\rangle_f$ and $|11\rangle_f$ devote themselves to the odd-FP current. However, due to $\Gamma_1\ll\Gamma_0$, the suppression of $I_J^{(e)}$ only appears in a narrow region near the point of $\varepsilon_0=0$, while the $2\pi$-period oscillation of $I_J^{(o)}$ distributes in a wide region accompanied by its enhanced amplitude [See Fig.4(a) and Fig.4(c)]. Next, in the presence of the intradot Coulomb interaction, $\varepsilon_s$ splits into two, i.e., $\varepsilon_s$ and $\varepsilon_s+U$. Accordingly, in the energy region of $-U<\varepsilon_s<0$, the fermion in the QD is changeable between 0 and 1, which magnifies the transformation of the Josephson effect caused by the shift of QD level. Since a finite magnetic field on the QD plays a similar role in affecting the fermion occupation in the QD, the magnetic field makes a similar contribution to the Josephson effect, compared with the Coulomb interaction. In addition, it should be noticed that the Fano interference induces the asymmetric spectra of the Josepshon currents vs $\varepsilon_0$.

\begin{figure}
\centering\scalebox{0.46}{\includegraphics{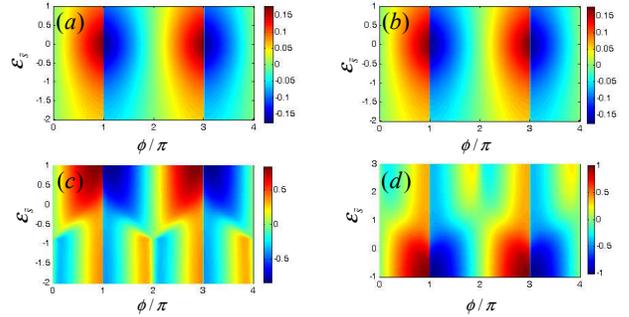}}
\caption{Josephson currents in the limit of strong magnetic field on the QD. The left and right columns describe the even and odd FP results. (a)-(b) The cases of $\Gamma_0=0$. (c)-(d) Results of $\Gamma_0=0.5$.} \label{Direct2}
\end{figure}
At last, we would like to pay attention to the extreme case of strong magnetic field where only one level (i.e., $E_{\tilde{s}}$) contributes to the Josephson effects. In such a case, the matrix dimension of ${\cal H}^{(e)}_T$ and ${\cal H}^{(o)}_T$ will be halved, as discussed in the above section. The corresponding numerical results are shown in Fig.6. First, in Fig.6(a)-(b) we can see that in the case of $\Gamma_0=0$, the Josephson currents in different FPs are the same as each other, with their $2\pi$ period. On the other hand, when the direct coupling between the Majorana doublets is considered, the Josephson currents become dependent on the FP. As shown in Fig.6(c), in the even-FP case, increasing $\varepsilon_{\tilde{s}}$ can change the current period from $\pi$ to $2\pi$ with the clear transition region near $\varepsilon_{\tilde{s}}\approx-1.0$. However, in the odd-FP case, similar result occurs when $\varepsilon_{\tilde{s}}$ decreases. These results can certainly be clarified by discussing the influence of the fermion number in the QD on the FP of states $|n_fn_{\tilde{f}}\rangle_f$. Also, in Fig.6(c)-(d) we can find that the Fano interference induces the dissimilar transition behaviors of the Josephson currents in different FPs.

 \section{summary\label{summary}}
In summary, we have investigated the Josephson effect in the Majorana-doublet-contributed junction modified by the different inter-doublet coupling manners. It has been found that an embedded QD in this junction plays a nontrivial role in modifying the Josephson effects, since the tunable fermion occupation in the QD re-regulates the FP of the Majorana doublets for conserving the FP in whole system. As a result, the $4\pi$-period, $2\pi$-period, and $\pi$-period Josephson currents have opportunities to come into being, respectively. To be concrete, when the Majorana doublets couple indirectly via a non-magnetic QD, the normal Josephson effects occur, and the FP change just causes the reversal of the current direction and the variation of the current amplitude. Only in the odd-FP case, can applying finite magnetic field on the QD induce the appearance of the fractional Josephson effect. When the direct and indirect couplings between the Majorana doublets coexist, no fractional Josephson effect takes place, regardless of finite magnetic field on the QD. Moreover, the $\pi$-period current has an opportunity to appear with the shift of the QD level. We believe that this work will be helpful for describing the QD-assisted Josephson effects between the Majorana doublets.

\par

This work was financially supported by the Fundamental Research
Funds for the Central Universities (Grants No. N130405009 and No. N130505001), the
Natural Science Foundation of Liaoning province of China (Grant No. 2013020030), and the Liaoning BaiQianWan Talents
Program (Grant No. 2012921078).

\clearpage

\bigskip

\end{document}